# Quantum Computations with Optical Waveguide Modes

Jian Fu

Department of Optical Electrical Engineering, Zhejiang University,
Hangzhou 310027, China

**ABSTRACT**

A fully optical method to perform any quantum computation with optical waveguide modes is proposed by supplying the prescriptions for a universal set of quantum gates. The proposal for quantum computation is based on implementing a quantum bit with two normal modes of multi-mode waveguides. The proposed universal set of gates has the potential of being more compact and easily realized than other optical implementations, since it is based on planar lightwave circuit technology and can be constructed by using Mach-Zehnder interferometer configurations having semiconductor optical amplifiers with very high refractive nonlinearity in its arms.

A great deal of effort has gone into the search for a practical architecture for quantum computation. Recently, the work has focused on NMR [1], solid-states [2], and atomic [3, 4], but so far none of these systems has demonstrated all of the desired features such as strong coherent interactions, low decoherence, and straightforward scalability. As was recognized early on, single-photon optics provides a nearly perfect arena for many quantum-information applications despite the absence of significant nonlinear effects of photon-photon interactions at the quantum level [5]. Schemes of optical quantum gates have been proposed in the last few years [6, 7]. Such models typically make use of the Kerr nonlinearity to produce intensity-dependent phase shifts, so that the presence of a photon in one path induces a phase shift to a second photon. Typical optical nonlinearities are so small that the dimensionless efficiency of photon-photon coupling rarely exceeds in orders of $10^{-10}$ [7]. Due to this weak coupling, it is much more difficult to construct a 2 qubit gate which operates at the single-photon level. A few schemes have been proposed for producing the enormous nonlinear optical responses necessary to perform quantum logic at the single-photon level. Such schemes involve coherent atomic (slow light [8] and E.I.T. [9]) or photon-exchange interactions [10]. Another scheme was proposed by Knill, Laflamme and Milburn or quantum computation based on linear optics [11], demonstrating that this obstacle can be overcome.

In recent years, planar lightwave circuit (PLC) technology has been rapidly developed to meet fiber communication systems required [12]. PLC technology is based on creating optical waveguides on substrates using manufacturing processes similar to semiconductors. An optical waveguide is a set of optically transparent layers which guide light within them. It is constructed by building these layers on top of a substrate material which provides physical support and a flat, pure layer to deposit on. The light is confined to the 'guiding' layer of relatively high refractive index (RI) surrounded above and below by lower index cladding materials. This confines the light vertically; horizontal control is provided by lithographically limiting the extent of guiding or cladding layers. Solving Maxwell's equations directly subject to the boundary conditions of the planar waveguide structure [12, 13], we can derive the possible solutions of

Maxwell's equations consisting of a discrete spectrum of a finite number of normal modes plus a continuum of waveguide (radiation) modes, that are spatial (transverse) modes of classical or quantum states of the optical field. All the normal modes, each of which is normalized and orthogonal to each of the others, constitute a complete set of solutions for Maxwell's equations in the sense that an arbitrary solution can be expanded in terms of them. An unperturbed waveguide can transmit any of its normal modes without converting energy to any of the other possible normal modes or to the continuous spectrum. But any slight perturbation of the guide, such as a series of waveguide transitions/junctions or two separate waveguides brought into proximity with each other, couples the particular normal mode to all other normal modes even to the modes of the continuum. When a resonance condition is satisfied, a slight perturbation of the waveguide can cause a large exchange of power between the modes of the unperturbed waveguide [13].

In this paper, we suggest using a set of discrete waveguide modes for implementing a QC. The fundamental units of QC are qubits, the quantum generalizations of classical bit. Qubits can be realized by two normal modes of dual-mode waveguides, such as the zero logical state $|0\rangle$ encoded into one normal mode and the logical one $|1\rangle$ given by other orthogonal normal mode. A qubit's state space consists of all superpositions of the basic normal modes $|0\rangle$ and $|1\rangle$. These qubits are not only applied to quantum states of optical field such as single photon states, but also to classical states such as coherent states. By using a multimode waveguide Mach-Zehnder interferometer (MZI), directional couplers (DCs) and other nonlinear optical devices such as semiconductor optical amplifiers (SOAs), we propose a fully optical method to perform quantum computation.

We now discuss the advantages of quantum computing with waveguide modes over other optical quantum computation. A SOA base on PLC technology is similar to a semiconductor laser diode, except that the reflectivity of the end faces is deliberately minimized to suppress lasing. Thus the SOA acts as a one-pass device for a lightwave with an inversion that is created by electrical pumping. The conduction and valence bands in a semiconductor can be modeled as an ensemble of two level atom-like systems. For photons that are resonant with the transition energy levels of the states that are inverted, stimulated emission can occur, i.e., photons at these frequencies obtain a gain. As the intensity of light increases, the gain saturates from the depopulation of the conduction band due to stimulated emission. Associated with this change in the gain due to saturation is a refractive index change, as described by the Kramer-Kronig dispersion relations. The refractive nonlinearity of SOAs is about $10^{-6} cm^2/W$ and $10^8$ times larger than an equivalent length of silica fiber. Optical switches using SOAs as the nonlinear switching element, have been used to demonstrate switching in systems using low control pulse energy (250 fJ) [14]. All-optical 3R (reamplified, reshaped and retimed) regeneration in optical communications systems along with wavelength conversion at 80 Gbit/s with error-free operation has been demonstrated using cross-phase modulation (XPM) in a nonlinear MZI with SOAs [15]. All-optical switches and wavelength-conversion devices based on XPM in SOAs using the MZI or Michaelson configuration have been integrated on PLC and are reviewed in [14, 16]. Lasers with quantum dot (QD) active material have shown record-low threshold current densities and it is of interest to explore the potential of QD SOAs to enhance the nonlinear interaction between two photons to implement quantum computation at the single-photon level

[17]. Moreover, optical waveguide can increase the non-linearity due to very high optical intensity in the core area. Therefore, QC based on PLC technology can be easily realized than other optical implementation. Eventually as processes for combining hybrid elements develop [18], it may be possible to have active and passive devices on one chip as well and thus the possibility of an Erbium-Doped Waveguide Amplifier or true loss-less components which include built-in amplification to compensate for insertion loss. So another important advantage for quantum computing with optical waveguide modes is that PLC technology allows a much tighter density of optical and electronic components given that all functions are performed on a single 'quantum CPU' chip.

Considering a simple three-layer waveguide structure and deriving a solution of Maxwell's equation for the guided modes of the structure, we obtain electric-field profiles as shown in Fig. 1. These Cartesian components of the transverse electric (TE) field are solutions of the scalar wave equation $\{\nabla_x^2 + \nabla_y^2 + k^2 n^2(x, y) - \beta^2\} \Psi = 0$, where $n(x, y)$ is the refractive-index profile, $k = 2\pi/\lambda$, $\lambda$ is the free-space wavelength. The solution $\Psi(x, y)$ of the scalar wave equation and its first derivatives are everywhere continuous and are therefore bounded. This leads to an eigenvalue equation for the allowed discrete values of $\beta$. The eigenfunctions with discrete eigenvalues are called the normal modes of the waveguide, which constitute a complete set of functions in the sense that an arbitrary solution of the scalar wave equation can be expanded in term of them. In Fig. 1, the geometry and optical wavelength are assumed such that the structure supports two normal modes, namely $TE_0$ mode and $TE_1$ mode.

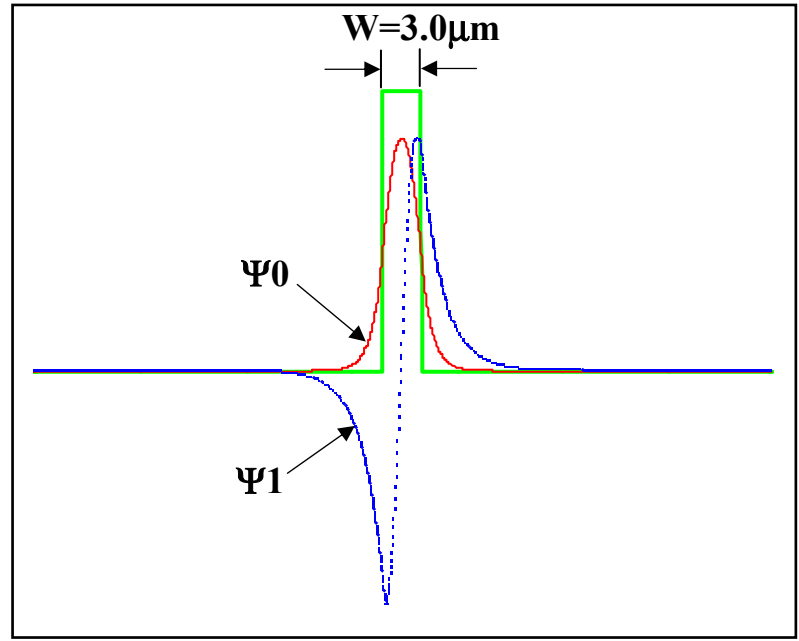

Figure 1: Electric field profiles of *x* or *y* for the first two normal modes $TE_0$ ($\Psi_0$) and $TE_1$ ($\Psi_1$) of a three-layer structure, and the RI distribution of the structure, RI $n_{core}$=1.57, width W=3.0μm for the core layer, and $n_{clad}$=1.55 for the cladding layer.

We denote the first mode as $\Psi_0(x,y)$ and the second mode as $\Psi_1(x,y)$ with propagation constants $\beta_0$ and $\beta_1$, respectively. As we know, if the profile $n(x,y)$ is independent of z, arbitrary local fields $\Psi(x,y,z)$ propagating in the waveguide at position z can be described by a superposition of two normal modes $\Psi_0(x,y)$ (the TE$_0$ mode) and $\Psi_1(x,y)$ (the TE$_1$ mode), that is, $\Psi(x,y,z) = C_0\Psi_0(x,y)e^{-i\beta_0 z} + C_1\Psi_1(x,y)e^{-i\beta_1 z}$, where $C_0$ and $C_1$ are the amplitudes of the modes $\Psi_0(x,y)$ and $\Psi_1(x,y)$. When a dual-mode waveguide has nonuniformities which vary distance z along its length, propagation $\Psi(x,y,z)$ can be described by a set of coupled equations based on the set of normal modes { $\Psi_0(x,y)$, $\Psi_1(x,y)$ }, that is, $\Psi(x,y,z) = C_0(z)\Psi_0(x,y)e^{-i\beta_0 z} + C_1(z)\Psi_1(x,y)e^{-i\beta_1 z}$, where $C_0(z)$ and $C_1(z)$ denote the couple-mode amplitudes. By using coupled mode theory [13], the coupled-mode equations for M coupled dual-mode waveguides are obtained,

$$\frac{dC_j^{(k)}(z)}{dz} = -\frac{i\omega}{4}\frac{\beta_j^{(k)}}{|\beta_j^{(k)}|}\sum_{k'}\sum_{l=0,1} K_{jl}^{kk'} C_l^{(k')}(z) e^{i\left(\beta_j^{(k)}-\beta_l^{(k')}\right)z} \quad (1)$$

where $K_{jl}^{kk'} = \int \Psi_j^{(k)*}\left[n^2(x,y,z)-n^2(x,y)\right]\Psi_l^{(k')}dxdy$, $n(x,y,z)$ is the z-dependent refractive-index profile, the superscript k denotes the kth waveguide and the subscript j denotes the jth order mode.

Given that we are using $|0\rangle$ (the TE$_0$ mode, $\Psi_0(x,y)$) and $|1\rangle$ (the TE$_1$ mode, $\Psi_1(x,y)$) to represent logical 0 and 1, respectively. First, let us consider a dual-mode waveguide MZI (shown in Fig. 2) with a phase shifter $\phi$ in one of its arms. The analysis of the device [13] consists in finding a unitary transformation connecting the input field $|\Psi_i\rangle = C_0^i|0\rangle + C_1^i|1\rangle$ and the output field $|\Psi_o\rangle = C_0^o|0\rangle + C_1^o|1\rangle$. We obtain the unitary transformation

$$U = \begin{pmatrix} \cos(\phi/2) & i\sin(\phi/2) \\ i\sin(\phi/2) & \cos(\phi/2) \end{pmatrix} \quad (2)$$

where the phase $\phi$ accounts for any phase shift between the two arms of the MZI, and $C_0^o = \cos\frac{\phi}{2}C_0^i + i\sin\frac{\phi}{2}C_1^i$, $C_1^o = \cos\frac{\phi}{2}C_1^i + i\sin\frac{\phi}{2}C_0^i$. The MZI can be configured in two extreme positions by choosing $\phi=0$ and $\phi=\pi$. In the former, all inputs are unchanged after the gate, while the latter acts as a quantum NOT gate. All inputs to $|0\rangle$ appear in the $|1\rangle$ output and vice versa, extra an additional phase. If choosing $\phi=\frac{\pi}{2}$, $|0\rangle \to \frac{1}{\sqrt{2}}(|0\rangle+i|1\rangle), |1\rangle \to \frac{1}{\sqrt{2}}(|1\rangle+i|0\rangle)$, which can be used to generate the desired superposition state. In order to confirm the validity of the present analysis, we have performed more rigorous numerical analysis using the effective index method [13] and the finite-difference beam propagation method (FD-BPM) [19]. The parameters used are, the RI of core and cladding of waveguides $n_{core} = 1.57$, $n_{clad} = 1.55$, respectively, the width of waveguide $W = 3.0 \mu m$, and the wavelength $\lambda = 1.064 \mu m$. Fig.3 illustrates the optical simulation of quantum NOT gate using a MZI whose arms have a phase difference caused by a phase shifter. When the phase shifter's length $L = 1mm$ and RI difference $\Delta n = 0.0008$, the input state $|0\rangle$ is transformed to the output state $|1\rangle$ and vice versa.

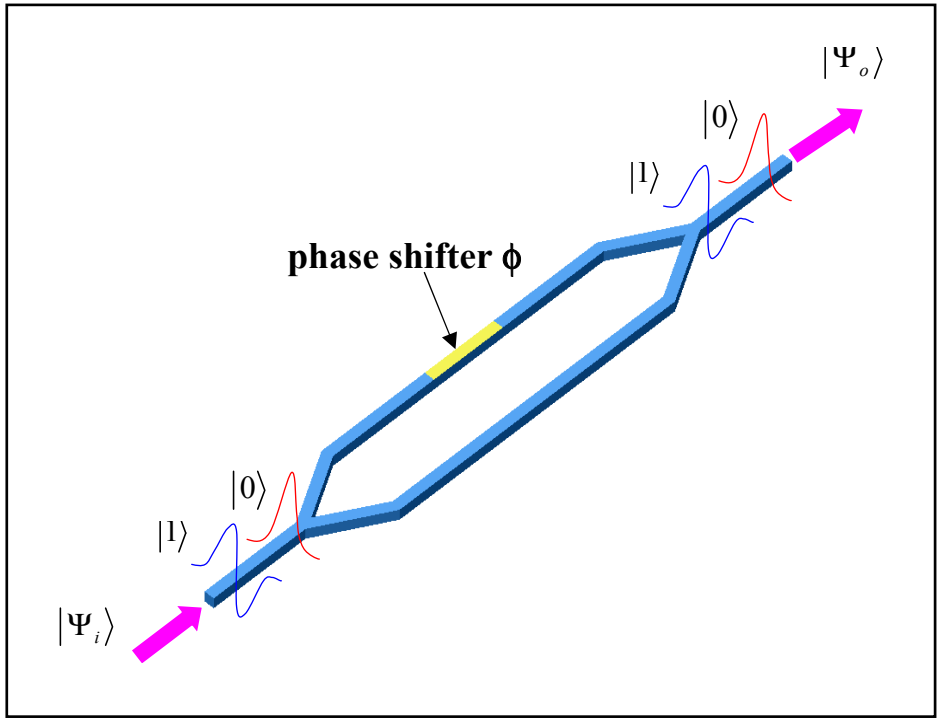

Figure 2: Geometry of the MZI whose arms have a phase difference ϕ caused by a phase shifter.

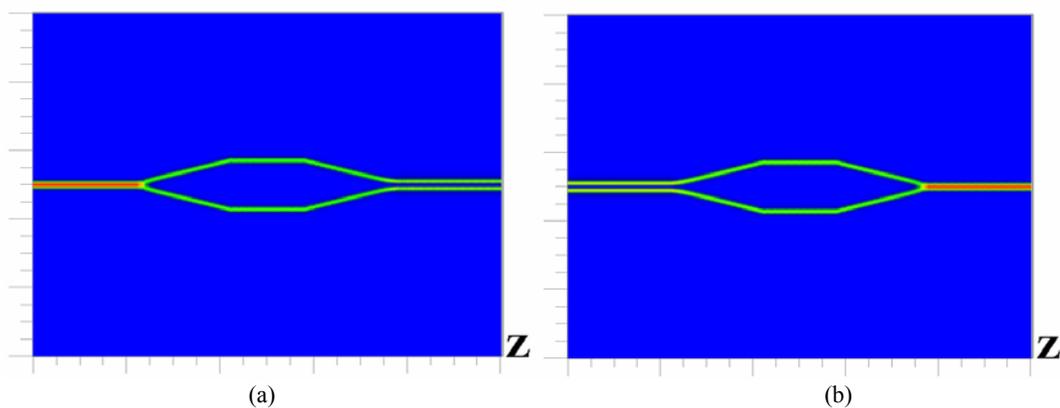

(a)           (b)

Figure 3: FD-BPM simulation results for quantum NOT gate using a MZI whose arms have a phase difference $\phi = \pi$ caused by a phase shifter, (a) $|0\rangle \rightarrow |1\rangle$, and (b) $|1\rangle \rightarrow |0\rangle$, where $z$ from 0 to 5000 $\mu$m.

Next, we consider a dual-mode waveguide DC with uniform coupler region of length *L*. Power transfer between the modes in two waveguides is described by the coupled-mode equations (1). After neglecting the weak coupling between different order modes, we obtain that the field amplitude in each of the two waveguides varies according to

$$\begin{cases} C_j^{(1)}(z) = \cos(\kappa_j z) C_j^{(1)}(0) - i \sin(\kappa_j z) C_j^{(2)}(0) \\ C_j^{(2)}(z) = -i \sin(\kappa_j z) C_j^{(1)}(0) + \cos(\kappa_j z) C_j^{(2)}(0) \end{cases} \quad (3)$$

where the coupling coefficients $\kappa_j = -\frac{\omega \varepsilon_0}{4} \int \Psi_j^{(2)*} \left[ n^2(x,y,z) - n^2(x,y) \right] \Psi_j^{(1)} dxdy$, the superscript 1, 2 denote the first and second waveguides respectively and the subscript $j = 0, 1$ denotes the *j*th order mode. Now we discuss a 'mode-separated/combined' device constructed from a DC. The schematic illustration of the devices is shown as coupler regions in Fig. 4. Assuming the input state $|0\rangle$ into waveguide 1 at $z = 0$, we obtain $C_0^{(1)}(0) = 1, C_1^{(1)}(0) = 0$, $C_0^{(2)}(0) = 0, C_1^{(2)}(0) = 0$. Setting the state $|0\rangle$ remaining in the same waveguide at $z = L$, we obtain $C_0^{(1)}(L) = 1, C_1^{(1)}(L) = 0$, $C_0^{(2)}(L) = 0, C_1^{(2)}(L) = 0$. Also assuming the input state $|1\rangle$ into waveguide 1 at $z = 0$, we obtain $C_0^{(1)}(0) = 0, C_1^{(1)}(0) = 1$, $C_0^{(2)}(0) = 0, C_1^{(2)}(0) = 0$. But the output state $|1\rangle$ appearing in the other waveguide at $z = L$, we obtain $C_0^{(1)}(L) = 0, C_1^{(1)}(L) = 0$, $C_0^{(2)}(L) = 0, C_1^{(2)}(L) = -i$. Therefore, we obtain an explicit condition $\cos(\kappa_0 L) = \sin(\kappa_1 L) = 1$, which can be satisfied by selecting and adjusting the coupling coefficients and length to perform 'mode-separate/combine'.

By using the dual-mode waveguide MZI, DC and Kerr-like mediums, an optical model for a 'quantum' C-NOT gate is indicated schematically in Fig. 4. Essentially it is a dual-mode MZI that a substance with an intensity-dependent RI (XPM via Kerr effect or SOAs) is placed in both arms. The device works as follows. The qubits propagating in waveguides 1 and 2 are pertained to the control qubit and the target qubit, respectively. When $|1\rangle$ is present at the control bit, the intensity of the qubit is coupled into one arm of the MZI by the first DC. If the input control bit contains a field just sufficient to cause a phase shift of *π*, the states of the target bit will be flipped, namely $|0\rangle \to |1\rangle$, $|1\rangle \to |0\rangle$. Then the control bit is coupled back waveguide 1 by the second DC and left unchanged. When $|0\rangle$ is present at the control bit, the intensity of the qubit is never coupled into the arm of the MZI. Therefore the control and target qubits are left unchanged. The C-NOT gate that is implemented by using the device was numerically calculated by an improved FD-BPM [20] to simulate the propagation of waves in a Kerr-like nonlinear waveguide, the result is shown in Fig. 5.

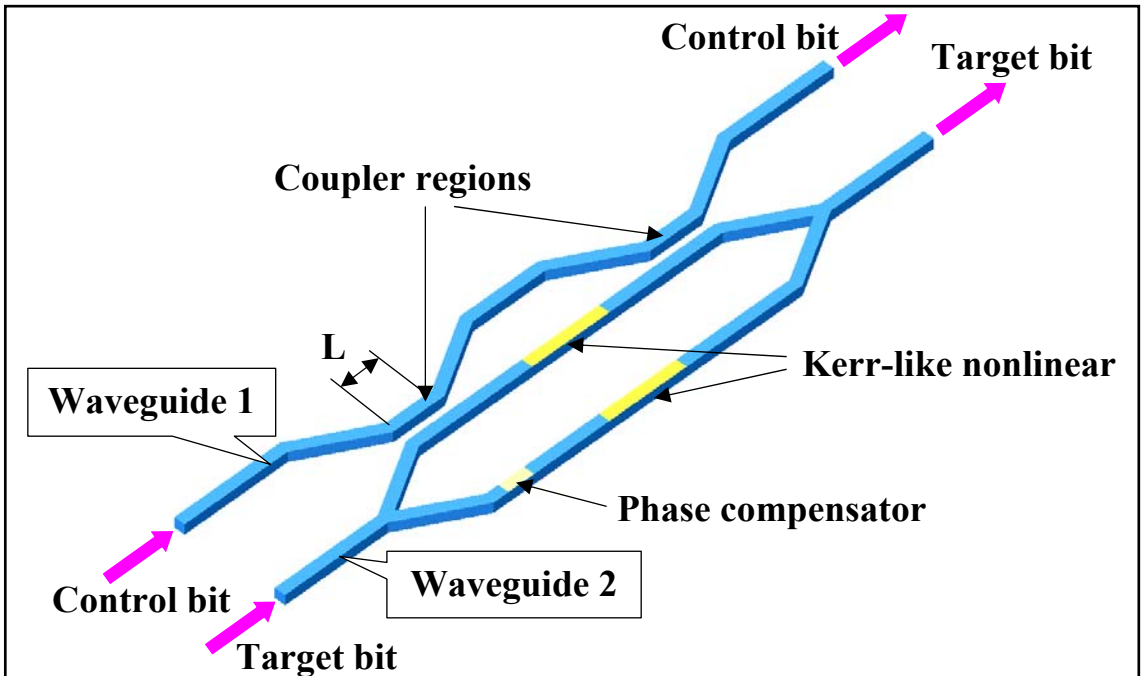

Figure 4: A 'quantum' C-NOT gate constructed using a nonlinear MZI (XPM via Kerr effect or SOAs) and two DCs as 'mode-separated/combined' devices. The coupler regions with interaction length $L = 823\mu m$, separation between waveguide cores $D = 1.2\mu m$ and the slope of transition region is 0.007.

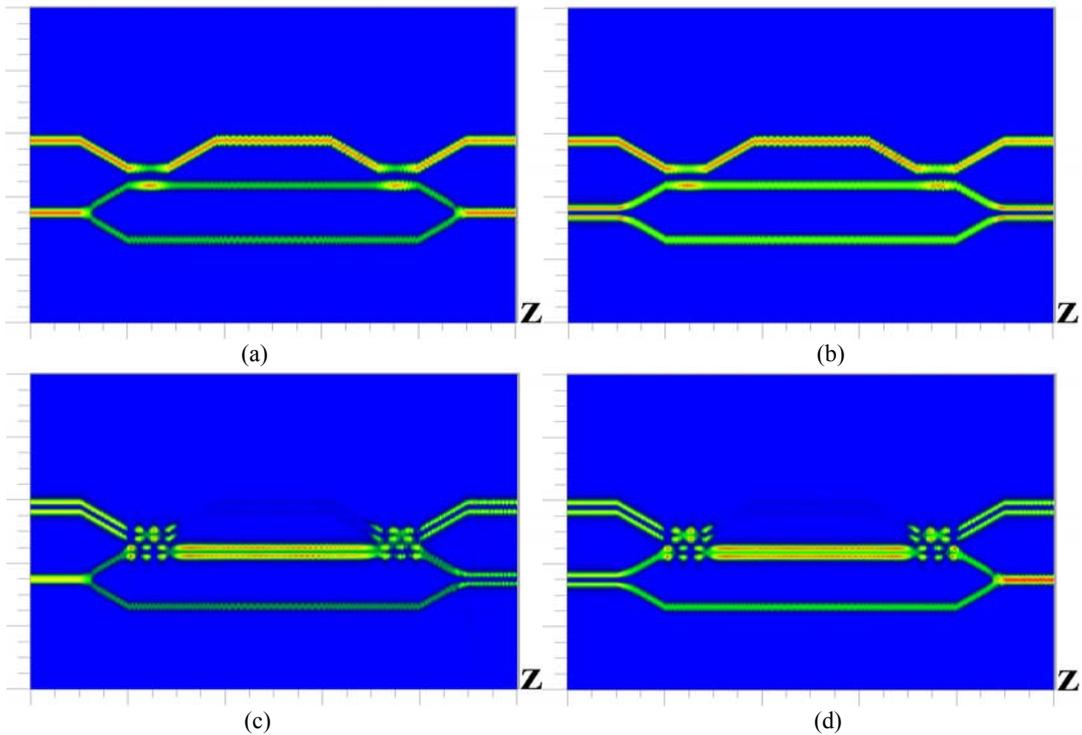

Figure 5: FD-BPM simulation results for the quantum C-NOT configuration shown in Fig. 4, (a) and (b) for the control bit $|0\rangle$, (c) and (d) for the control bit $|1\rangle$, where z from 0 to 10000 $\mu$ m.

The measurements of the output states, to be performed as the final step of a quantum computation, consist of mode separated or cut-off devices and optical receivers. These may be easily achieved by means of the 'mode-separated' DC or a single mode waveguide to cut-off higher order modes, followed by PIN/APD optical receivers.

Two important imperfections which lead to quantum computation errors are energy loss and decoherence. The former occurs due to absorption in waveguide media and radiation loss caused by waveguide bends [13] or sidewall imperfections [21] *etc*. Any energy losses on the control bit will cause phase errors in one arm of the MZI. Based on built-in amplification technology, this problem can be remedied. But decoherence is present even in cases in which energy loss is negligible. In the QC with waveguide modes, the major sources of decoherence include both the photons interacting each other through a Kerr-like medium, and any imperfection of the waveguides, such as deviation from perfect waveguide straightness or a local change of waveguide's RI, *etc*. The latter is a unique source of decoherence for the optical QC we are analyzing. On second thoughts, we consider that sidewall roughness of waveguides is one of the most important factors to cause decoherence and energy loss. According to [21], the imperfections of the waveguide wall transfer energy from one guided mode to other guided modes, which cause mode disorder, namely decoherence, and the radiation field of the continuum of unguided modes, which cause energy loss. Typical roughness rms values for waveguides fabricated by conventional photolithography and reactive ion etching techniques are about 10 nm [22]. The sidewall roughness is greatly improved from 10nm rms to 2nm rms or less through improved etching and smoothing processes [22, 23]. By using the method in [21], we roughly estimate that a length of 10% power transfer (decoherence) from mode $TE_0$ to $TE_1$ in a silica waveguide with $\Delta = 1\%$ index difference is more than 100cm. The waveguide sidewall smoothing technology [24] could lead to significant improvements in practical waveguide design for optical quantum computing devices.

Whether quantum computation with optical waveguide modes can be implemented in practice remains to be proved by experiments. However, the results obtained here have shown that, in principle, the present scheme may open new perspectives for practical quantum computation.